# Advances in Bioinformatics and Computational Biology: Don't take them too seriously anyway.

Emanuel Diamant
VIDIA-mant, Kiriat Ono, Israel

**Abstract -** *In the last few decades or so, we witness a paradigm shift in our nature studies – from a data-processing based computational approach to an information-processing based cognitive approach. The process is restricted and often misguided by the lack of a clear understanding about what information is and how it should be treated in research applications (in general) and in biological studies (in particular). The paper intend to provide some remedies for this bizarre situation.*

## 1   Introduction

Striking advances in high-throughput sequencing technologies have resulted in a tremendous increase in the amounts of data related to various biological screening experiments. Consequently, that gave rise to an urgent need of new techniques and algorithms for analyzing, modeling and interpreting these huge amounts of data.

To reach this goal, Computational Biology and Bioinformatics techniques and tools are being devised, developed and introduced into research practice.

What is the difference between the two? Wikipedia does not see any difference at all [1]. NIH working definition, [2], distinguishes only a slight disparity between them:

"Computational biology uses mathematical and computational approaches" (to reach its goals), while "Bioinformatics applies principles of information sciences and technologies" (for the same purposes).

Perhaps the most evident difference lies in their historical background. Computational biology starts when the "brain as a computer" metaphor becomes generally accepted as the dominant research paradigm. Therefore, almost all scientific fields have become "computational" – Computational neuroscience, Computational genomics, Computational chemistry, Computational ecology, Computational linguistics, Computational intelligence, and so on. It was acknowledged then that the surrounding world is represented by data that is sensed by our sensors and thus processing of this data (making computation on it) was accepted as the prime duty of the research community.

At the same time, it was acknowledged that human interaction with the external world can be seen as a communication process by which sensory data is delivered to the conscious mind. For such a case, Shannon's "Mathematical Theory of Communication", [3], and the Information Theory embedded in it have been developed and become the dominant research paradigm of the second half of the past century. Obviously, this was the ground on which Bioinformatics has emerged and has gained its recognition as a separate research field.

However, Shannon's information is restricted only to data communication issues. Message meaning (semantics) – a crucially important part of a communication process – is totally omitted from its considerations. That explains the visible similarity between Computational Biology and Bioinformatics – both are first of all busy with data processing, at the same time, both are deficient in dealing with information issues (due to the lack of understanding about the essence of information).

The intention of this paper is to attempt to clarify the existing confusion.

## 2   So, what is information?

A proper definition of the term "information" does not exist. Therefore, I would like to propose my own one. It is an extended version of the Kolmogorov's mid-60s definition [4], which can be now expressed in the following way:

**"Information is a linguistic description of structures observable in a given data set".**

A digital image would serve us as a testbed for definition analysis. An image is a two-dimensional set of data elements called pixels. In an image, pixels are distributed not randomly, but due to the similarity in their physical properties, they are naturally grouped into some clusters or clumps. I propose to call these clusters **primary or physical data structures**.

In the eyes of an external observer, the primary data structures are further arranged into more larger and complex assemblies (usually called "visual objects"), which I propose to call **secondary data structures**. These secondary structures reflect human observer's view on the primary data structures composition, and therefore they could be called **meaningful or semantic data structures**. While formation of primary data structures is guided by objective (natural, physical) properties of the data, ensuing formation of secondary structures is a subjective process guided by human habits and customs.

As it was already said, **Description of structures observable in a data set should be called "Information".** In this regard, two types of information must be distinguished – **Physical Information and Semantic Information**. They are both language-based descriptions; however, physical information can be described with a variety of languages (recall that ma-thematics is also a language), while semantic information can be described only by means of the natural human language. (More details on the subject can be find in [5]).

Every information description is a top-down evolving coarse-to-fine hierarchy of descriptions representing various levels of description complexity (various levels of description details). Physical information hierarchy is located at the lowest level of the semantic hierarchy. The process of sensor data interpretation is reified as a process of physical information extraction from the input data, followed by an attempt to associate the physical information at the input with physical information already retained at the lowest level of a semantic hierarchy. If such association is reached, the input physical information becomes related (via the physical information retained in the system) with a relevant linguistic term, with a word that places the physical information in the context of a phrase, which provides the semantic interpretation of it. In such a way, the input physical information becomes named with an appropriate linguistic label and framed into a suitable linguistic phrase (and further – in a story, a tale, a narrative), which provides the desired meaning for the input physical information.

## 3  New wine in old wineskins

In the light of the above elucidation, the mutual interrelations between Computational Biology and Bioinformatics can be now explained and put into action: Essentially, Computational Biology is an attempt to mimic physical information descriptions while Bioinformatics is an attempt to mimic semantic information descriptions. Now, all further advances in their development have to take into account the integration-dissociation peculiarities and task division strategy following from the new information definition.

Let me put it again: semantic perception of the sensed data begins with physical information extraction from it. It must be emphasized that only physical information is being processed further in the semantic information-processing stream. All physical traits of the input data are lost at this stage. In the end, we understand the essence of an image ignoring its illumination conditions or color palette. The same is with speech perception – we understand the meaning of a phrase independent of its volume or gender voice differences.

The extracted physical information is associated then with the physical information retained at the lowest level of the semantic hierarchy. In such a way, it finds its place in a linguistic expression, which determines its meaning, its semantics. (Analogous to "comprehension from usage" or "understanding from action" forms of semantics disambiguating).

This physical data structures naming is in a close resemblance to the ontology-based annotation process. Ontologies are the most recent form of knowledge representation and are widely used in biomedical science enabling to turn data into knowledge. Despite of the resemblance, semantic information hierarchies and ontologies are strikingly different. From my definition of semantic information follows that 1) knowledge is memorized (retained in the system) information (and nothing else!), 2) semantic information is an observer's property, and 3) semantic information has nothing to do with data! That is, data is semantics devoid. So, the purpose of ontologies "to describe the semantics of data", is misinterpreted. Computational biology tools developers have to pay more attention to this peculiarity.

## 4  Conclusions

One can hardly overestimate the importance of physical and semantic information segregation. For the first time, data-based information and its semantic (language-based) interpretation are detached and now can be treated correctly and essentially.

For the first time, information is represented as a linguistic description, as a string of words, a piece of text. It does not matter that in biotic applications these texts are written in the four-letter nucleotide alphabet. The important thing is that now information is **materialized**, and as such can be stored, retrieved, changed, transmitted and (generally speaking) processed as any other material object.

In this regard, the paradigm shift from data-based computational approach to information-based cognitive approach receives its proper theoretical underpinning, which will certainly promote its further development and utilization.

One of the obvious problems that arises in such a transition is as follows: We are accustomed to use computers in our everyday life. Computer is a data processing devise. Semantic information comes about as a text string. Therefore, semantic information processing must be treated as text strings processing. But that is not what our computers are supposed to do. There is an urgent need to invent a new generation of computers that will be capable to process natural language texts (which are the expression of semantic information).